\documentclass[a4paper,twocolumn]{article}
\usepackage{hyperref}

\usepackage[dvips]{graphicx}

\begin{document}

\twocolumn[

\title{Laser Doppler Imaging of Microflow}

\author{M. Atlan$^1$, M. Gross$^1$ and J. Leng$^{2}$}

\date{}

\maketitle

 $^1$Laboratoire Kastler-Brossel, Ecole Normale Sup\'{e}rieure, UMR 8552 (ENS, CNRS, UMPC), 24 rue Lhomond 75231 Paris cedex 05. France  \\

$^2$Laboratoire Microfluidique, MEMS, Nanostructures, ESPCI, CNRS UPR A0005,  Universit\'{e} Pierre et Marie Curie, 10 rue Vauquelin 75005 Paris. France.\\

%

\begin{abstract}
We report a pilot study with a wide-field laser Doppler detection scheme used to perform laser Doppler anemometry and imaging of particle seeded
microflow. The optical field carrying the local scatterers (particles) dynamic state, as a consequence of momentum transfer at each
scattering event, is analyzed in the temporal frequencies domain. The setup is based on heterodyne digital holography, which is used to map
the scattered field in the object plane at a tunable frequency with a multipixel detector. We show that wide-field heterodyne laser Doppler
imaging can be used for quantitative microflow diagnosis; in the presented study, maps of the first-order moment of the Doppler frequency
shift are used as a quantitative and directional estimator of the Doppler signature of particles velocity.
\end{abstract}

\bigskip


\bigskip

]


\bigskip

\textbf{Citation:}

M. Atlan, M. Gross and J. Leng. "Laser Doppler imaging of microflow"
Journal of the European Optical Society - Rapid publications, \textbf{1}, 06025 (2006)
 
 [DOI: 10.2971/jeos.2006.06025]

\section{Introduction: Velocimetry and microfluidics}

Microfluidics has emerged as an important tool for engineering,
fundamental science, biology, etc. [1]. The ability to control
flow patterns at a typical scale of a micrometer through
advanced microfabrication techniques [2] offers a neat and
powerful guide for the development of analytical microfluidics
chips. Of primary importance is the charaterization of
flow patterns in microsystems [3] and much of the technological
effort has focused on direct determination of flow profiles
of liquids inside micro-capillaries. The size reduction indeed
precludes the use of external probes and therefore dismisses
most of traditional macroscopic techniques for velocimetry
(eg. hot-wire anemometry [4]). Particle-based flow visualization
techniques (particle image velocimetry (PIV) [5], that often
use fluorescent microspheres of a size of order of 100 nm
in dilute suspensions at volume fraction of order of 10.5) are
popular as little intrusive tools that offer high spatial resolution
in the characterization of local flow properties (especially
ì-PIV, for biological or complex fluids [6][8] and biphasic
flows [9]). Holographic configurations have been used for
particle field diagnosis [10] and velocimetry [11, 12]. Digital
holographic PIV (HPIV) constitutes an active field of research,
where off-axis HPIV appears to be a better configuration than
inline HPIV in terms of signal-to-noise ratio (SNR) [13, 14].
Other visualisation techniques, based on fluorescence bleaching
[15], correlation spectroscopy [16], caged fluorescence [17],
molecular tagging velocimetry [18], total internal reflection
fluorescence [19, 20], etc., are also used to investigate local
flow properties.

In conventional laser Doppler velocimetry (LDV), the scattered
laser light coming from a focus point onto a sample is
detected by a single detector and analyzed by a spectrum analyzer.
The power spectrum of coherent monochromatic light
scattered by moving particles is broadened as a result of momentum
transfer. The resulting broadening of light is linked to
the velocity distribution of scatterers [21, 22]. Doppler maps
can be realized by scanning the focus point on the sample
surface, which constitutes the principle of scanning Laser
Doppler Imaging (LDI) [23][25], for which spatial resolution
is typically low. Although LDV is a standard technique
for macroflow studies, several difficulties limit its use in microflow:
The focusing point of the laser beam has to be very
accurately controlled, its micron-scale extension limits the accuracy
of the velocity measurements, and the temporal resolution
diminishes with the spatial scale of the sample [3, 26].

We designed a parallel imager aimed at LDI [27], alleviating
the issue of spatial scanning. Our technique involves a frequency
selective heterodyne detection of the light on a multi
pixel CCD detector which records digital holograms in an off axis
configuration. The image is reconstructed numerically by
using standard numerical holography algorithms. By sweeping
the optical frequency of heterodyne detection local oscillator
(reference arm), we record the spatial distribution of tunable
frequency components of the object field sequentially.We
demonstrate here that the technique can be employed as a tool
for in-plane flow analysis in microfluidic networks and is especially
well-suited for building up Doppler maps (absolute
values along with gradient measurements) on intermediate
scale ($\sim$ 1 cm) microfluidics webs.

Our technique can be interpreted as a Doppler global (or planar)
velocimetry (DGV) measurement, but on a very different
temporal frequency scale than molecular absorption-based
DGV [28, 29]. Molecular absorption DGV relies on the absorption
characteristics of iodine vapor to convert a Doppler
shift to a recordable intensity. Because of the characteristics
of the absorption curve of the iodine and the finite stability
of laser cavities, the frequency resolution is $\sim$ 1 MHz (velocity
resolution of $\sim$ 1 m/s) [30], and is therefore not suitable
for microflow analysis. In our case, the measurement relies
on a heterodyne detection extremely selective in frequency.
the detection bandwidth is the reciprocal of the measurement
time, hence a frequency resolution of a few Hz, compatible
with microflow analysis. Our method also differentiates itself
from scanning LDI and HPIV methods. Whereas scanning
LDI performs sequential measurements in space and parallel
measurements in the temporal frequency domain, our instrument
does parallel measurements in space and sequential
measurements in frequency. And unlike HPIV and its refinements
(e.g. [31]), our technique does not rely on the localization
of the seeding particles onto an image. Like in DGV,
Doppler maps are the result of a direct selective measurement
of frequency-shifted components of the scattered light. The
advantages of our heterodyne holography method for wide
field laser Doppler measurements are the sensitivity (heterodyne
gain), the selectivity in the temporal frequency domain
(coherent detection) and the large optical \'{e}tendue (product of
the detector area by the detection solid angle) of multi pixel
coherent detection [32, 33].

\section{Optical Setup}

\begin{center}
\begin{figure}[h]
  \includegraphics[width=8 cm]{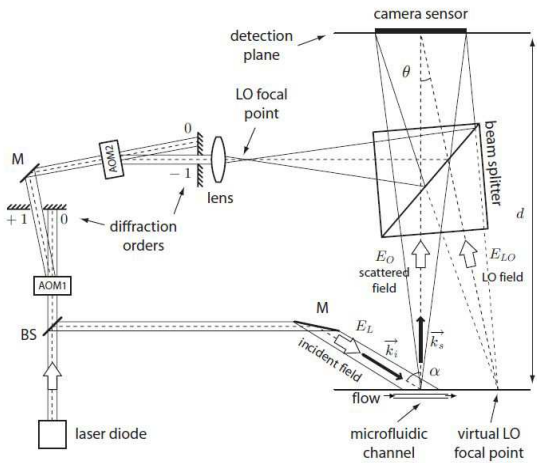}\\
  \caption{Optical configuration : lensless holographic setup. EL: laser (incident) field. EO:
field scattered by the object. $E_{LO}$: reference (local oscillator) field. $\textbf{k}_i$: incident wave
vector. $\textbf{k}_s$: scattered wave vector, in the direction of the receiver. AOM : acousto-optic
modulator. BS : beam splitter. M : mirror.
}\label{fig1}
\end{figure}
\end{center}

The experimental setup [27, 34] is based on an optical interferometer
sketched in Figure 1.A CW, 80mW, $\lambda_0$ = 658 nm diode
(Mitsubishi ML120G21) provides the main laser beam, split
by a prism into a reference (local oscillator, LO) and an object
arm. The dimensionless scalar amplitudes of the fields are
noted $E_L$ (incident field), $E_O$ (scattered, object field) and $E_{LO}$
(LO field). The object is illuminated by the laser beam with an
incidence angle a. Scattered light is mixed with the reference
beam and detected by a PCO PixelFly 1.3 Mpix CCD camera
(1280 $\times$ 1024 square pixels, pixel size: 6.7 µm, framerate
$\omega/2\pi$ = 4 Hz, exposure time $\tau_E$ = 125 ms), set at a distance $ d
= 50$ cm from the object. Two Bragg cells (acousto-optic modulators,
AOM, Crystal Technology) are used to shift the LO by
a tunable frequency (the difference of their driving frequencies).
A 10 mm focal length lens is placed in the reference arm
in order to create an off-axis ($\theta \approx 1 ^\circ$ tilt angle) virtual point
source in the object plane (see Figure 1). This configuration
constitutes a lensless Fourier holography setup [35].

To perform a heterodyne detection of a frequency component
of the object field, the LO field frequency $\omega_{LO}$ is detuned by
$\Delta \omega - \omega_S /n$ with respect to the main laser beam frequency $\omega_L$.
The $\Delta \omega$ shift allows the LO field to match the part of the object
field shifted by $\Delta \omega$. The additional shift at the n-submultiple
of the camera sampling frequency provokes a modulation of
the interference pattern sampled by the detector at $\omega_S /n$.

The measured power spectrum S associated to the field EO is [27]:
\begin{equation}\label{Eq1}
    S(\Delta \omega) = A ~|E_O(\omega_L + \Delta \omega)|^2
\end{equation}
Where A is a positive constant. The Dw spectral point is readout
from a sequence of n images recorded by the camera
[27]. EO is recorded in the detector plane by the frequencyshifting
technique [32] which is the dynamic equivalent of the
phase-shifting digital holography technique [36]. The lensless
Fourier holographic setup [35] used here is less demanding
in numerical calculations than the general Fresnel holography
configuration. The numerical reconstruction algorithm of the
image is limited to one discrete Fourier transform [37, 38].
Artefacts due to the finite size of the sensor and the spatial
discrete Fourier transform are negligible [39] (narrower than
one pixel in the reconstructed image).

We define the signal $S_{dB}$ represented on spectra profiles and
maps:
\begin{equation}\label{Eq2}
    S_{dB}(\Delta \omega) = 10\log_{10} \left[  \frac{S(\Delta \omega)}{N(\Delta \omega)}\right]
\end{equation}
where$ N(\Delta \omega) $ is the quantity $S(\Delta \omega)) $ assessed in a region of the
reconstructed hologram where the object light contribution is
null. It was reported to be shot-noise dominated [33].

To map the projection of velocities with the in-plane component
of the incident field wavevector (according to Eq. 4),
we compute the first moment of the frequency shift $\Delta f =
\Delta \omega /(2\pi)$:
\begin{equation}\label{Eq3}
   \langle \Delta f \rangle= \frac{1}{2\pi} \frac{\sum \Delta \omega S(\Delta \omega)}{\sum S(\Delta \omega)}
\end{equation}
where $\sum S(\Delta \omega)$ is the zeroth-order moment. These sums are
calculated over the full range of measured shifts.

\section{Microfluidic devices with velocity gradients}

\begin{center}
\begin{figure}[h]
  \includegraphics[width=8 cm]{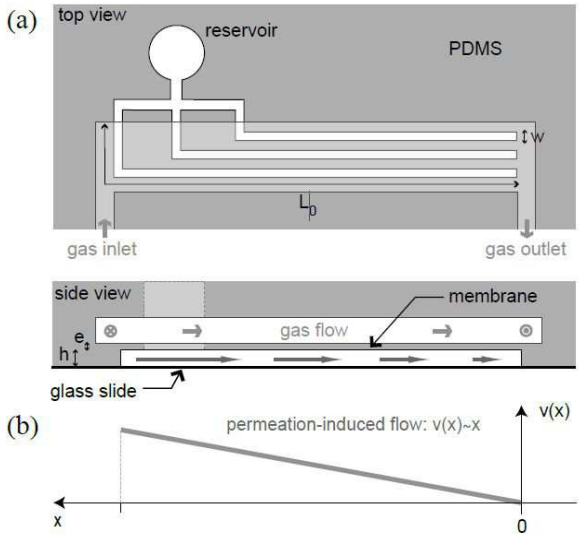}\\
  \caption{ a. Sketch of the microfluidic device generating linear velocity fields via evaporation
[40]. This devices is made of PDMS on glass and has two layers: one for the fluid
separated from the other for the gas removal by a thin PDMS membrane ($ e \sim  10 \mu m$).
Typical dimensions are $L_0 \sim 1 cm$, $h \sim 20 \mu m$. b. The permeation-induced flow
is stationnary, and varies linearly in space between 0 and a typical upper bound
$v_{max} \approx  50 \mu /s $ (longest channel) which actually depends on $L_0$.
}\label{fig2}
\end{figure}
\end{center}

In this study, we designed a fluidic network made using standard
microfabrication techniques [2]: a circuit is patterned
onto a silicon substrate using micro-photolithography (spincoating
of a photoresist exposed to UVs through a mask
and developped to obtain hard patterns resolved at $ \approx 1 \mu m$),
molded into a transparent elastomer poly-dimethylsiloxane
(PDMS) and stuck on a glass substrate. We thus obtain closed
circuits for liquid transport of typical dimensions 10 × 50 ×
5000 $\mu m$. The actual devices we use are special (Figure \ref{fig2}) as
they generate a flow via evaporation in a dead end microchannel
[40], with a flow field which varies in space. Details for
fabrication of these 2-layer systems (one layer for liquid transport,
one layer for gas transport for evaporation) are given
elsewhere [41] and the main feature we shall retain is the linear
dependence of mean velocity in a channel as a function of
position (while by design [40], these devices are used to concentrate
a solute near the dead end of the channels, an effect
we shall also evidence here). Such a heterogeneous flow field
is a specificity of permeation-induced flows [42], and we will
exploit it for testing the construction of velocity maps.

The microsystems are filled with a solution of pure water
seeded with 1.0 ìm diameter latex spheres (with carboxylate
stabilizing surface groups, by Molecular Probes). Evaporation
induces a permanent flow that drives the liquid from
the reservoir towards the dead end of the channel, with a velocity
slowing down linearly from $\approx 50 \mu $m/s down to 0 (see
Figure 2). This result holds for the mean velocity; two effects alter
this simple view: the flow field actually shows a parabolic
profile (Poiseuille flow) and the particles undergo Brownian
dynamics.

\section{Wide field Laser Doppler mapping of microfluidic webs}

\begin{center}
\begin{figure}[]
  \includegraphics[width=8 cm]{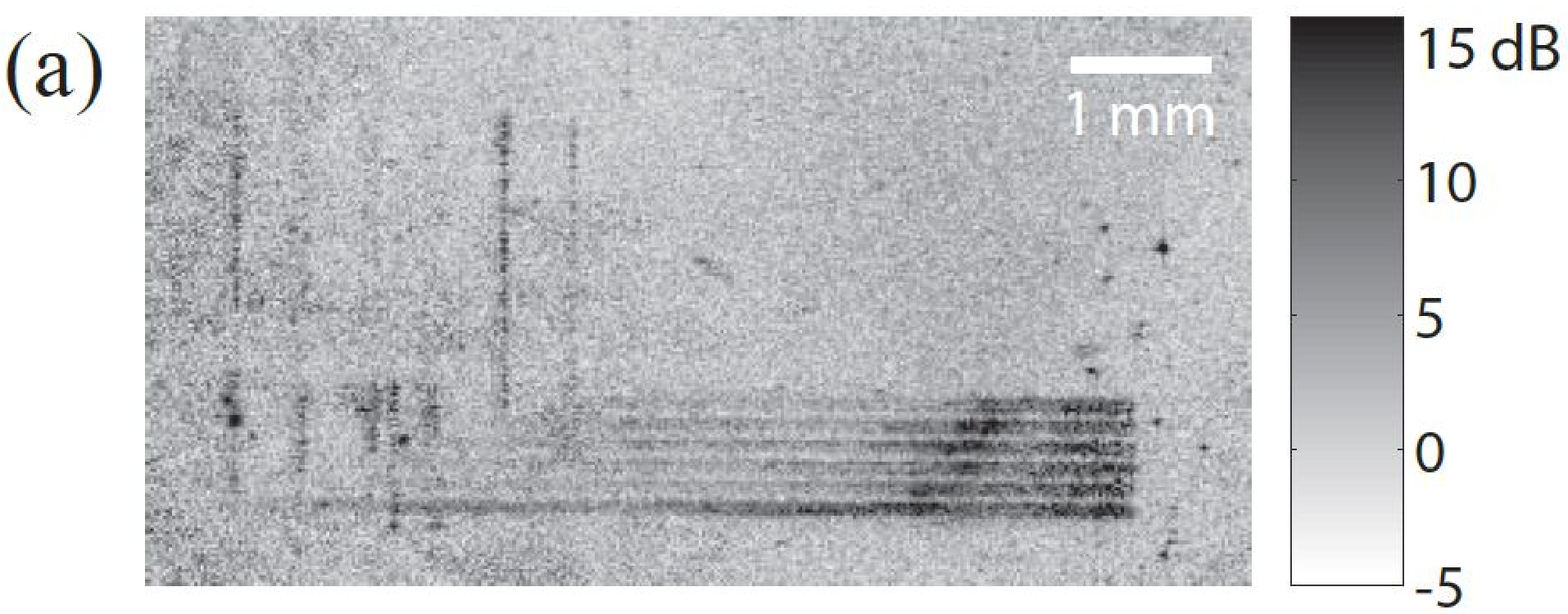}\\
 \includegraphics[width=8 cm]{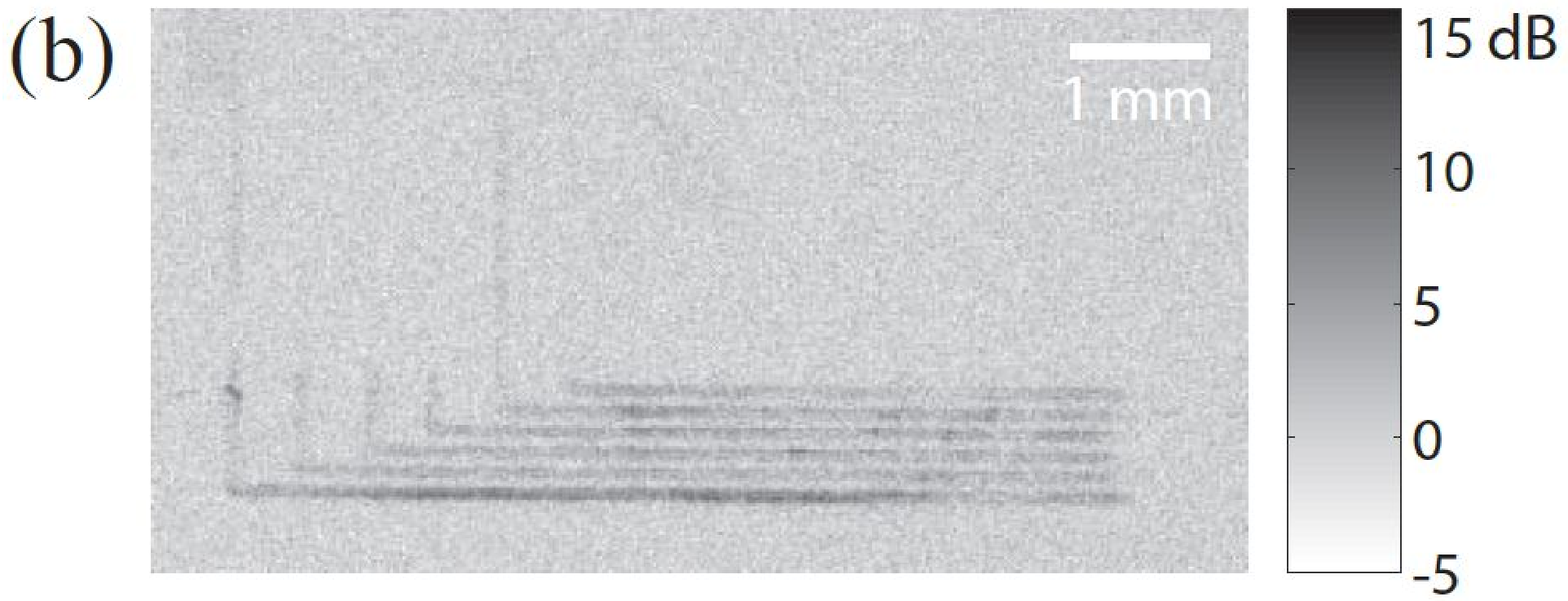}\\
 \includegraphics[width=8 cm]{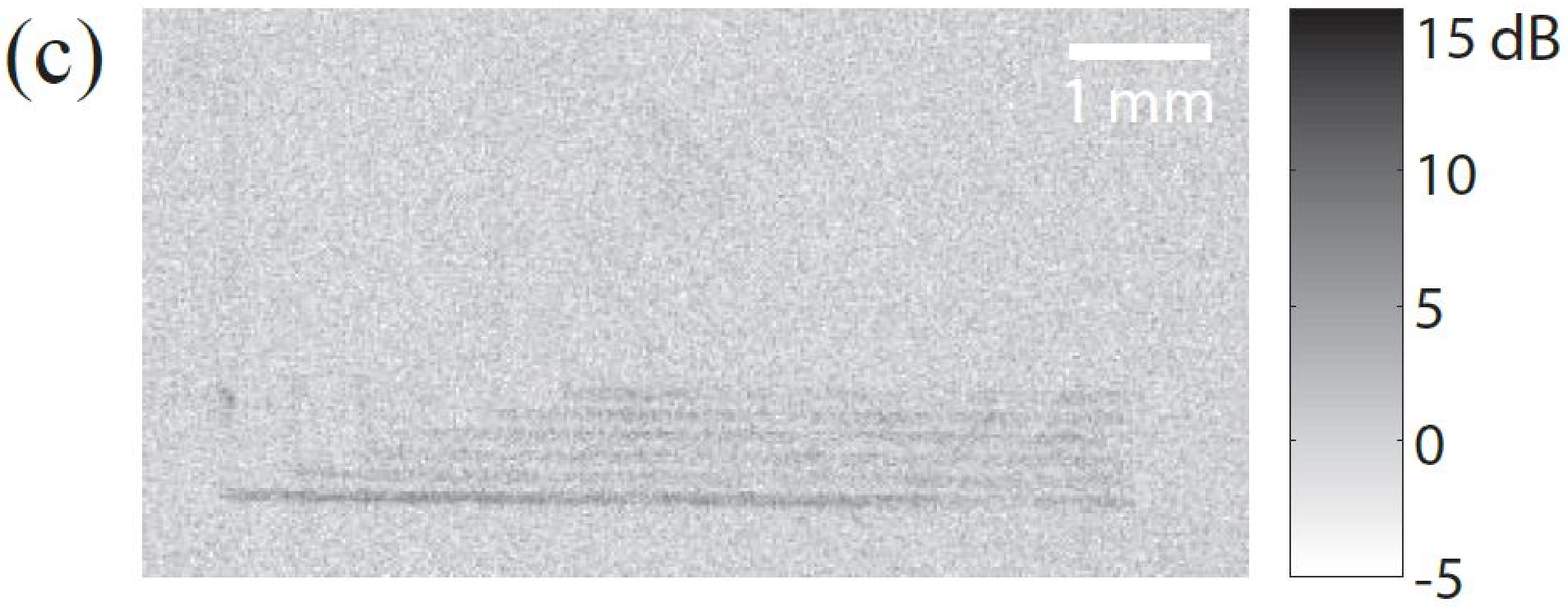}\\
 \includegraphics[width=8 cm]{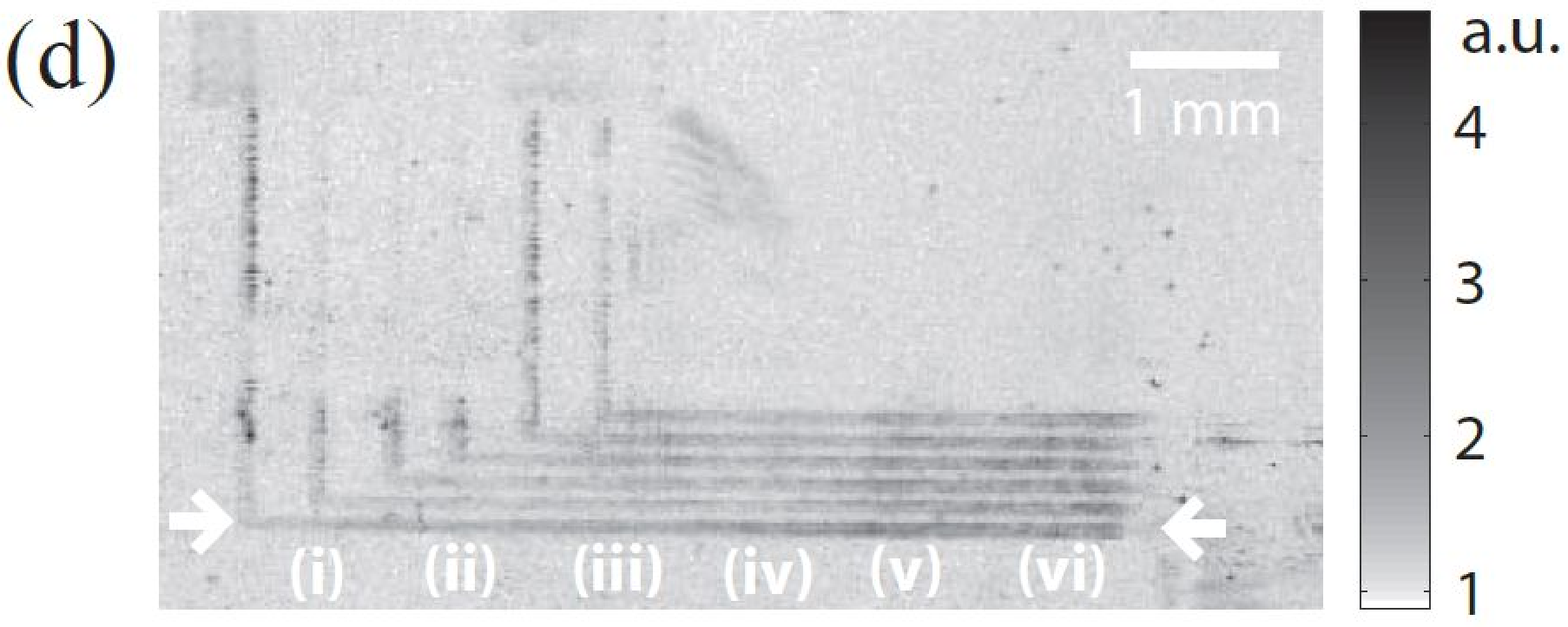}\\
  \caption{ (a) to (c) : $S_{dB}$ maps at three frequency shifts.
  $\Delta f$ = 0 Hz (a), $\Delta f$  = .22 Hz
(b), $\Delta f$  = .44 Hz (c). (d) : Zeroth-order moment of the frequency shift, arbitrary
units. Linear scale. Order of magnitude of velocities : 10 $\mu$m/s. accumulation of $m$ =
16 images, $\omega_S/(2_pi)$ = 4 Hz, $n$ = 8-phase demodulation.
}\label{fig3}
\end{figure}
\end{center}

\begin{center}
\begin{figure}[]
  \includegraphics[width=8 cm]{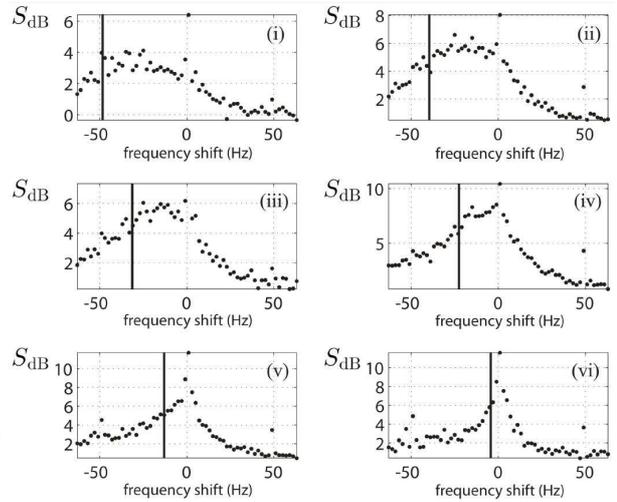}\\
  \caption{ Spectra averaged over 50 $\times $ 4 pixels in the six ROIs outlined in Figure3; $S_{dB}$
vs. $\Delta f$. Vertical markers represent the expected average flow-induced frequency shift $\langle \omega_D \rangle$ in each region of interest, for which neither brownian motion nor static scattering
are taken into account.
}\label{fig4}
\end{figure}
\end{center}

\begin{center}
\begin{figure}[]
  \includegraphics[width=8 cm]{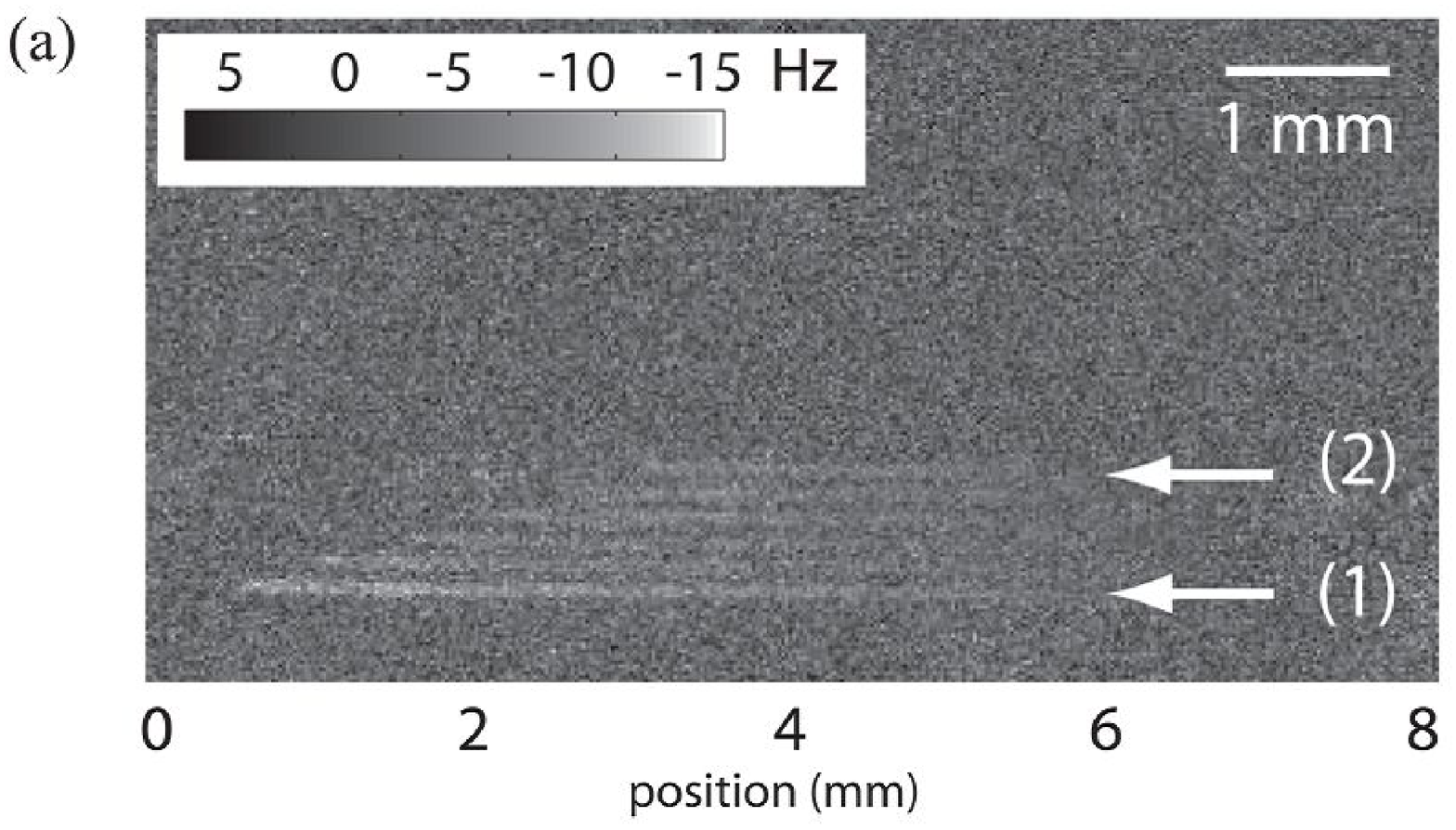}\\
  \includegraphics[width=8 cm]{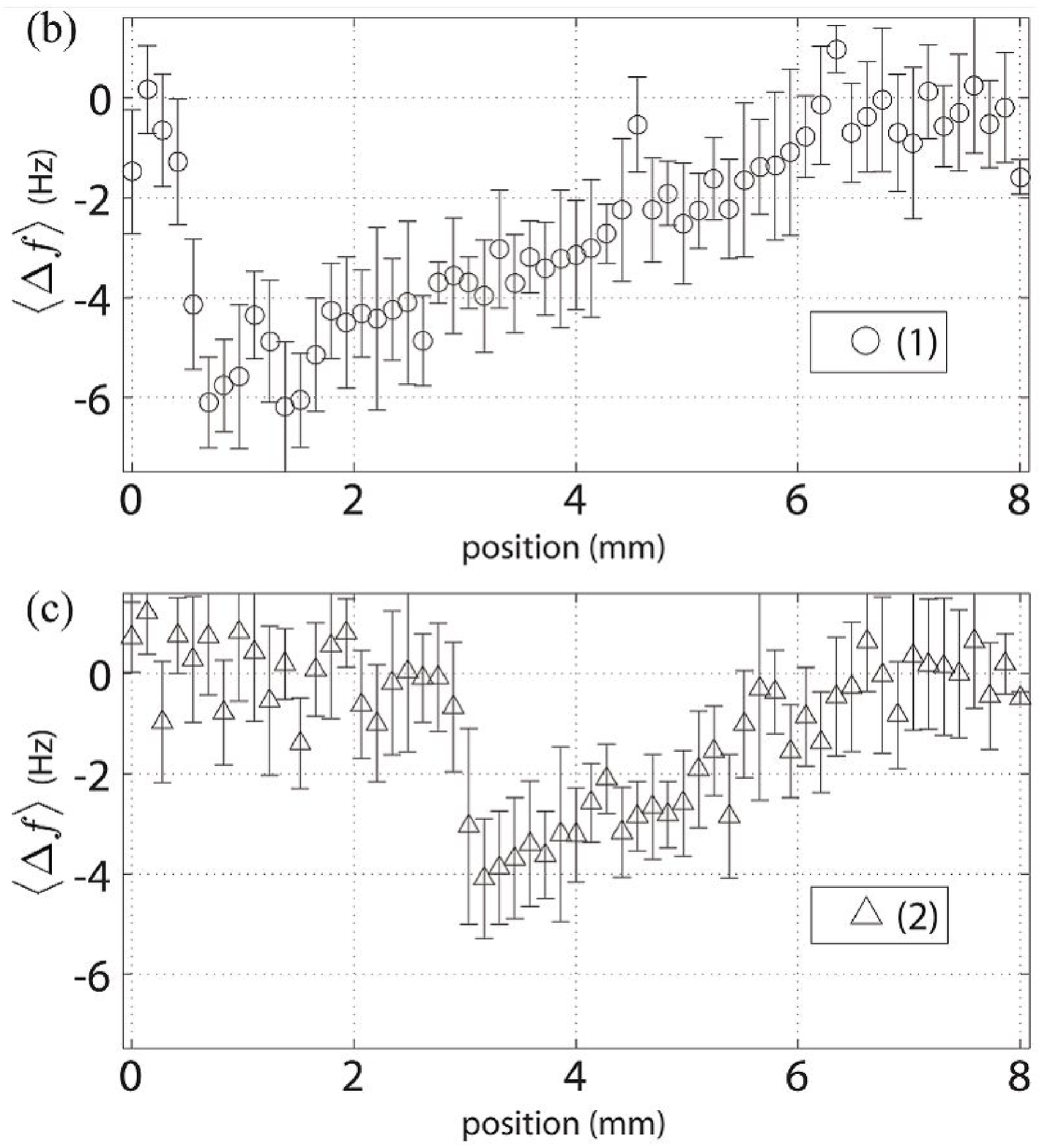}\\
  \caption{(a) : First moment of the frequency shift, in Hertz. (b) and (c: same quantity
averaged over $7 \times 4$ pixels, and plotted versus longitudinal position in canals 1 and
2, outlined in Figure 5a
}\label{fig5}
\end{figure}
\end{center}

A $2 \times 10^{-4}$ volumic concentration suspension of latex beads
is perfused through the microfluidic web. This concentration
of latex corresponds to $3.82\times 10^{-4}$ particles per cubic micron,
which was the lowest seeding for which those particles would
lead to an acceptable SNR, in our configuration. A Mie algorithm \footnote{from C. Maetzler, iapmw.unibe.ch}
was used to assess its optical parameters : the resultant
mean free path is $l_s = 1.5 $mm, and the anisotropy coefficient
is $g = 0.93$. This suspension of latex beads was perfused
in the microfluidic device, composed of 6 channels of
$h = 23 \mu \textrm{m} × w = 100 \mu \textrm{m}$  cross section, which optical thickness (defined as the thickness of the channel in mean free path
units) $h \times \mu s \sim 10-2$ is much smaller than unity. The incidence
angle a of the laser beam (see Figure 1) is $\sim 45^\circ $. In this experiment,
Doppler shift maps were measured in the -64 Hz
$\rightarrow$ +64 Hz range (spacing interval : 2 Hz). $S(\Delta \omega)$ was calculated
with a $n$ image demodulation and averaged over a $m$
image sequence. Here : $m = 16$, $n = 8$. The total measurement
time of the spectral cube took $\sim$ 8 minutes and 30 seconds
(= 16 images $\times$ 65 frequency shifts $\times$ 1/4 second exposure
time). Nevertheless, in these conditions, the measurement of
one complete spectral map at an arbitrary frequency shift $ \Delta f$
took only 16 × 1/4 = 4 seconds.

Three $S_{dB}$ maps, measured at $\Delta f$ = 0, -22 and -44 Hz are represented
on Figures 3a to c. The whole microfluidic web appears
to respond at each arbitrary  $\Delta f$ frequency. Such maps
are difficult to interpret, since they represent a signal which
combines both scatterers density and Doppler signature. Additionally,
these results reveal an averaged response over the
height of the channel. The zeroth-order moment map of the
frequency shift, represented on Figure 3d, displays a signal
allegedly proportional to the moving particles concentration
[43] (under the approximation of a small scatterers densities),
which shows the expected increase in density at the end of the
channels, in region (vi).

The Doppler shift for one scattering event by a particle in
translation with velocity v is [44] :
\begin{equation}\label{Eq4}
    \omega_D= \textbf{q.v}
\end{equation}
where $\textbf{q} = \textbf{k}_s - \textbf{k}_i$ is the momentum transfer, $\textbf{k}_i$ is the incident
wave vector, ks the scattered wave vector and v the scatterer
instant velocity. In the chosen optical configuration (Figure
1), $\textbf{k}_s \textbf{· v}$ = 0. Hence the Doppler shift in a region where
the average speed of the particles is $\langle v \rangle$ :
\begin{equation}\label{Eq5}
   \langle  \omega_D \rangle = - \frac{2\pi}{\lambda_0} \sin(\alpha) \langle v \rangle
\end{equation}
At the entrance of the longest canal (canal (1) in Figure 5a)
the particles translation speed is $v \sim 50 \mu \textrm{m}.s^{-1}$. According to
the linear behaviour [40] of $v(x)$, the expected flow-induced
Doppler shift $ \langle \omega_D \rangle /(2\pi)$ for each region of interest (ROI) defined
in Figure 3b is represented by a vertical marker in Figures
4i to vi, at values ranging linearly from 49.2 Hz (i) to 4.4
Hz (vi), superimposed on each corresponding $S_{dB}$ spectrum.

In diluted samples (in very good approximation the ones
which optical thickness is much lower than 1), the root mean
square (RMS) Doppler shift of light scattered by a particle undergoing
brownian motion is $ \langle \omega^2_B \rangle ^{1/2} = \langle\textbf{ q}^2 \rangle D_B$  where$ D_B $is
the spatial diffusion coefficient of the brownian particle; in the
particular case of a spherical particle of radius $r$, its expression
is : $D_B = (k_B T)/(6 \pi \eta r)$, where $k_B$ is the Boltzmann constant
and $T$ the absolute temperature. For our latex beads in suspension
in water (of viscosity $\eta = 10^{-3}$ Pa.s), at room temperature
($k_B T = 4.0 \times 10^{-21}$ J), we have $D_B = 4.2 \times 10^{-13} \textrm{m}^2 \textrm{s}^{-1}$. The
average value of $\textbf{q}^2$, for a scattering angle ($\textbf{k}_i, \textbf{k}_s) = \beta$ is [45]:
\begin{equation}\label{Eq6}
    \langle \textbf{q}^2 \rangle =  4 \textbf{k}_i^2 \sin^2(\beta /2)
\end{equation}
which gives, in the chosen optical configuration (Figure 1):
\begin{equation}\label{Eq7}
    \langle \omega_B^2 \rangle^{1/2}=  \frac{16 \pi^2 D_B \sin^2((\pi - \alpha')/2)}{\lambda^2}
\end{equation}
where  $\alpha'$  is the incidence angle of light in water of refractive
index $n = 1.33$, and $ \lambda = \lambda_0/n$ is the optical wavelength in water.
The numerical value of $ \langle \omega_B^2 \rangle^{1/2}$  is
$\langle \omega_B^2 \rangle^{1/2}/(2\pi)=30.9$ Hz.

This value of the RMS brownian motion-induced Doppler
shift is compatible with the width observed, in Figure 4vi,
where the fluid velocity is nearly zero. In other regions : (i)
to (v), the observed spectrum corresponds to the convolution
of the brownian spectrum with the spectrum associated to the distribution of velocities in the flow (Poiseuille flow). The narrow
peak, observed on the spectra near zero frequency, is a
parasitic signal not related to the flow (light scattered by the
PDMS substrate...).

Figure 5 represents the first moment of the Doppler shift with
respect to the measured Doppler linewidths on each pixel,
mapped and plotted versus the longitudinal position along
two microfluidic channels. Figure 5b and 5c show a linear decrease
of $\langle \delta f \rangle$  with the position along channels (1) and (2),
highlighted in Figure 5a. This corresponds to the expected linear
decrease of the average velocity along the channels.

\section{Conclusion}

We have presented a wide field laser Doppler measurement
in a microfluidic device, and demonstrated that the velocity
resolution is compatible with microflow diagnosis. In the reported
pilot study, the setup, based on off-axis heterodyne
holography, performs a parallel measurement of the power
spectrum of the light scattered by a microfluidic device on a
CCD camera. The scheme is different from conventional laser
Doppler time-domain measurements : each frequency point of
the spectrum is acquired at a time, by tuning the LO frequency
at the desired shift. Since the spectrum is processed optically,
by an interferometric scheme between scattered light and a
detuned local oscillator, the span of the measurable spectrum
is not limited by the detector bandwidth. Furthermore, the optical
configuration allows to measure algebraic Doppler shifts
since the hologram records the optical field in quadrature,
rather than its intensity.

In this paper, the stress was put on the potential of the wide
field laser Doppler instrument in terms of frequency (and
hence velocity) diagnosis : high frequency resolution, high
SNR, high dynamic range of restituable frequencies on the
same image, and directional discrimination. The scope of the
presented study was to provide measurements in large scale
microfluidic devices into which large velocity gradients can
be created at the same time.

The spatial resolution of the digital holography-based apparatus
is about the extent of 1 pixel on the reconstructed image
[37]. Much better resolution (at the expense of the field of
view) can be obtained by using an holographic microscopy
configuration, which enables theoretically sub-micronic lateral
resolution [46] (up to the diffraction limit). Nevertheless,
time-averaged measurements of rarefied seeds at low spatial
scales, might become noisy and reduce the spatial resolution
of velocity maps.

The frequency resolution is given by the heterodyne bandwidth
of the detection (the inverse of the total acquisition
time of the image sequence). Although measured spectra are
stained by the Doppler linewidth contribution of Brownian
movement, the setup has an intrinsic ability to map small velocities
(up to $\sim$ 10 micron per second, but the actual limit
should be even lower, thanks to the heterodyne selectivity).
Since these (and lower) flow velocities are commonly encountered
in microfluidic webs, our frequency-domain wide-field
laser Doppler imager appears to be a valuable candidate for
microflow visualization and analysis.

The authors acknowledge support from the French ANR.

\bibliographystyle{unsrt}

\bibliography{reference}

[1] H. A. Stone, A. D. Stroock, and A. Ajdari, Engineering flows in
small devices: Microfluidics toward a lab-on-a-chip Annual Review
Of Fluid Mechanics 36, 381411 (2004).

[2] G. M. Whitesides, E. Ostuni, S. Takayama, X. Y. Jiang, and D. E.
Ingber,  Soft lithography in biology and biochemistry Annual
Review Of Biomedical Engineering 3, 335373 (2001).

[3] D. Sinton, Microscale flow visualization Microfluidics and
Nanofluidics 1, 221 (2004).

[4] G. Comte-Bellot, Hot-wire anemometry Annual Review Of Fluid
Mechanics 8, 209231 (1976).

[5] S. Kurada, G. W. Rankin, and K. Sridhar, Particle-imaging techniques
for quantitative flow visualization - a review Optics And
Laser Technology 25(4), 219233 (1993).

[6] H. Muller-mohnssen, D. Weiss, and A. Tippe, Concentration dependent
changes of apparent slip in polymer-solution flow Journal
Of Rheology 34(2), 223244 (1990).

[7] J. G. Santiago, S. T. Wereley, C. D. Meinhart, D. J. Beebe, and R. J.
Adrian, A particle image velocimetry system for microfluidics
Experiments In Fluids 25(4), 316319 (1998).

[8] A. F. Mendez-Sanchez, J. Perez-Gonzalez, L. de Vargas, J. R.
Castrejon-Pita, A. A. Castrejon-Pita, and G. Huelsz, Particle image
velocimetry of the unstable capillary flow of a micellar solution
Journal Of Rheology 47(6), 14551466 (2003).

[9] F. Innings and C. Tragardh, Visualization of the drop deformation
and break-up process in a high pressure homogenizer Chemical
Engineering \& Technology 28(8), 882891 (2005).

[10] BJ Thompson, JH Ward, and WR Zinky, Application of holographic
techniques for particle sizing analysis Applied Optics 6(3), 519
(1967).

[11] JD Trolinger, RA Belz, and WM Farmer, Holographic techniques
for the study of dynamic particle fields Applied Optics 8(5), 957
(1969).

[12] LM Weinstein, GB Beeler, and AM Lindemann, High-speed holocinematographic
velocimeter for studying turbulent flow control
physics American Institute of Aeronautics and Astronautics AIAA-
526 (1985).

[13] H. Meng, W.L. Anderson, F. Hussain, and D. D. Liu, Intrinsic
speckle noise in in-line particle holography Optical Society of
America Journal A 10, 20462058 (1993).

[14] H Meng and F Hussain, In-line recording and off-axis viewing
technique for holographic particle velocimetry Applied Optics 34,
1827 (1995).

[15] B. P. Mosier, J. I. Molho, and Santiago J. G, Photobleachedfluorescence
imaging of microflows Experiments In Fluids 33(4),
545554 (2002).

[16] C. Ybert, F. Nadal, R. Salome, F. Argoul, and L. Bourdieu, Electrically
induced microflows probed by fluorescence correlation spectroscopy
European Physical Journal E 16(3), 259266 (2005).

[17] W. R. Lempert, K. Magee, P. Ronney, K. R. Gee, and R. P. Haugland,
Flow tagging velocimetry in incompressible-flow using photoactivated
nonintrusive tracking of molecular-motion (phantomm)
Experiments In Fluids 18(4), 249 (1995).

[18] B. Stier and M. M. Koochesfahani, Molecular tagging velocimetry
(mtv) measurements in gas phase flows Experiments in Fluids
V26(4), 297304 (1999).

[19] C. Zettner and M. Yoda, Particle velocity field measurements in a
near-wall flow using evanescent wave illumination Experiments
in Fluids V34(1), 115121 (2003).

[20] K. D. Kihm, A. Banerjee, C. K. Choi, and T. Takagi, Near-wall
hindered brownian diffusion of nanoparticles examined by threedimensional
ratiometric total internal reflection fluorescence microscopy
(3-d r-tirfm) Experiments in Fluids V37(6), 811824
(2004).

[21] R. Bonner and R. Nossal, Model for laser doppler measurements
of blood flow in tissue Applied Optics 20, 20972107 (1981).

[22] J. D. Briers, Laser doppler and time-varying speckle: a reconciliation
Optical Society of America Journal A 13, 345 (1996).

[23] H Nakase, OS Kempski, A Heimann, T Takeshima, and J. Tintera,
Microcirculation after cerebral venous occlusions as assessed by
laser doppler scanning J. Neurosurg. 87(2), 307314 (1997).

[24] Beau M. Ances, Joel H. Greenberg, and John A. Detre, Laser
doppler imaging of activation-flow coupling in the rat somatosensory
cortex NeuroImage, 10(6), 716723 (1999).
[25] Ralf Steinmeier, Imre Bondar, Christian Bauhuf, and Rudolf
Fahlbusch, Laser doppler flowmetry mapping of cerebrocortical
microflow: Characteristics and limitations NeuroImage 15(1), 107
119 (2002).

[26] C.D. Meinhart, S.T. Wereley, and J.G. Santiago, Micron-Resolution
Velocimetry Techniques, in Developments in Laser Techniques and
Applications to Fluid Mechanics, (Springer, 1998).

[27] M. Atlan and M. Gross, Laser doppler imaging, revisited Review
of Scientific Instruments 77(11), 2006.

[28] H. Komine and S. J. Brosnan, Instantaneous, three-component,
doppler global velocimetry pages 273277 (1991).

[29] J. Meyers and H. Komine, Doppler global velocimetry: A new way
to look at velocity Laser Anemometry 1, 289 (1991).

[30] James F Meyers, Joseph W Lee, and Richard J Schwartz, Characterization
of measurement error sources in doppler global velocimetry
Measurement Science and Technology 12(4), 357368
(2001).

[31] C.-T Yang and H.-S Chuang, Measurement of a microchamber flow
by using a hybrid multiplexing holographic velocimetry Experiments
in Fluids V39(2), 385396 (2005).

[32] F. LeClerc, L. Collot, and M. Gross, Numerical heterodyne holography
with two-dimensional photodetector arrays Optics Letters
25(10), 716718 (2000).

[33] M. Gross, P. Goy, and M. Al-Koussa, Shot-noise detection of
ultrasound-tagged photons in ultrasound-modulated optical imaging
Optics Letters 28, 24822484 (2003).

[34] M. Atlan, M. Gross, T. Vitalis, A. Rancillac, B. C. Forget, and A. K.
Dunn, Frequency-domain, wide-field laser doppler in vivo imaging
Optics Letters 31(18) (2006).

[35] George W. Stroke, Lensless fourier-transform method for optical
holography Applied Physics Letters 6(10), 201203 (1965).

[36] U. Schnars and W. Juptner, Direct recording of holograms by a ccd
target and numerical reconstruction Applied Optics 33, 179181
(1994).

[37] Christoph Wagner, Sonke Seebacher, Wolfgang Osten, and Werner
Juptner, Digital recording and numerical reconstruction of lensless
fourier holograms in optical metrology Applied Optics 38,
48124820 (1999).

[38] U. Schnars and W. P. O. Juptner, Digital recording and numerical
reconstruction of holograms Meas. Sci. Technol. 13, R85R101
(2002).

[39] Thomas M. Kreis, Frequency analysis of digital holography Optical
Engineering 41(4), 771778 (2002).

[40] J. Leng, B. Lonetti, P. Tabeling, M. Joanicot, and A. Ajdari, Microevaporators
for the kinetic inspection of phase diagrams Physical
Review Letters 96(8), 084503 (2006).

[41] J. Goulpeau, D. Trouchet, A. Ajdari, and P. Tabeling, Experimental
study and modeling of polydimethylsiloxane peristaltic micropumps
Journal of Applied Physics 98(4) (2005).

[42] E. Verneuil, A. Buguin, and P. Silberzan, Permeation-induced
flows: Consequences for silicone-based microfluidics Europhysics
Letters 68(3), 412418 (2004).

[43] A. Serov, W. Steenbergen, and F. de Mul, Laser doppler perfusion
imaging with complementary metal oxide semiconductor image
sensor Optics Letters 27, 300 (2002).

[44] Y. Yeh and H. Z. Cummins, Localized fluid flow measurements
with an he-ne laser spectrometer Appl. Phys. Lett. 4, 176179
(1964).

[45] B. J. Berne and R. Pecora, Dynamic Light Scattering, (Dover, 2000).

[46] Ichirou Yamaguchi, Jun ichi Kato, Sohgo Ohta, and Jun Mizuno,
Image formation in phase-shifting digital holography and applications
to microscopy Applied Optics 40(34), 61776186 (2001).
06025-

\end{document}